\documentclass[fleqn,12pt,twoside]{article}
\usepackage{espcrc1,epsfig}

\usepackage{graphicx}
\usepackage[figuresright]{rotating}

\newcommand{\AmS}{{\protect\the\textfont2
  A\kern-.1667em\lower.5ex\hbox{M}\kern-.125emS}}

\title{Remarks on the extraction of freeze-out parameters}

\author{D.H. Rischke\address{RIKEN-BNL Research Center and Nuclear
        Theory Group, \\Brookhaven National Laboratory, 
        Upton, NY 11973, U.S.A.}%
        \thanks{Permanent address since February 2001:
        Institut f\"ur Theoretische Physik,
        Johann Wolfgang Goethe -- Universit\"at Frankfurt,
        Robert-Mayer-Str.\ 8--10, D-60054 Frankfurt/Main, Germany}}
       
\begin{document}

% typeset front matter
\maketitle

\begin{abstract}
I review the extraction of kinetic and chemical freeze-out parameters
from experimental data, with particular emphasis on the underlying assumptions
and the validity of the conclusions.
\end{abstract}

\section{INTRODUCTION}

Collisions of elementary particles, hadrons, 
and nuclei at ultrarelativistic energies produce a 
multitude of particles (``multiparticle production'').
Final-state interactions between the produced particles 
determine the dynamical evolution of the system.
In $e^+e^-$ and hadron-hadron collisions only few particles are produced, 
and it is unlikely that many final-state interactions occur. The
particles decouple (``freeze out'') from the system soon
after production.

On the other hand, in $AA$ collisions the density of produced particles
is sufficiently large over an extended
region in space-time, such that
the mean free path of produced particles becomes small and many final-state
interactions occur. These interactions drive
the system towards local {\em thermodynamic}, {\it i.e.},
{\em thermal}, {\em mechanical}, and {\em chemical\/} equilibrium.
In local thermodynamic equilibrium, the evolution of the system
is governed by the equations of ideal fluid dynamics \cite{dhrreview}. If
the system is only thermally and mechanically, {\it i.e.},
{\em kinetically}, but not chemically equilibrated,
these equations have to be supplemented by rate equations which
determine the chemical composition of the system \cite{dmedhr}.
In both cases, pressure gradients between dense, equilibrated matter 
and the vacuum drive collective expansion, which cools and dilutes 
the system. Freeze-out of particles occurs when microscopic interaction 
rates become smaller than the macroscopic expansion rate of the system.

By definition, after freeze-out the momenta of the produced particles 
do not change. The experimentally measured spectra of hadronic 
particles thus reflect the state of the system {\em at\/} freeze-out.
The question is whether the spectra can also tell us
about the state of the system {\em prior\/} to freeze-out?
For instance, can they tell us whether the system was in
thermodynamic, or at least kinetic equilibrium? Do they provide
information as to whether a quark-gluon
plasma (QGP), {\it i.e.}, an equilibrated state of quarks and gluons,
was created at some stage during the evolution of the system?

These questions are addressed in the following.
In section \ref{II}, I discuss whether the exponential nature of
single-inclusive, invariant particle spectra is sufficient to conclude
that the system was in thermal equilibrium. In section \ref{III},
it is argued that in $AA$ collisions collective expansion
is likely to occur prior to kinetic freeze-out.
In section \ref{IV}, I discuss how final-state particle ratios
yield information about the thermodynamic conditions at 
chemical freeze-out. Conclusions are given in section \ref{V}.

\section{EXPONENTIAL PARTICLE SPECTRA} \label{II}

In multiparticle production processes,
single-inclusive, invariant particle spectra
are typically exponential in the transverse momentum, $p_T$.
The reason for the exponential behavior is the phase space of
the final multiparticle state \cite{hagedorn}.
Consider for instance a process
where two particles (or nuclei) collide and produce $N$ particles
in the final state. Label the 4-momenta of the two incoming particles
as ${p_1'}^\mu$ and ${p_2'}^\mu$, 
and the 4-momenta of the $N$ outgoing particles
as $p_1^\mu,\, p_2^\mu,\, \ldots,\, p_N^\mu$. All 4-momenta are on-shell,
$p_i^\mu \equiv (E_i,{\bf p}_i)$, $E_i \equiv \sqrt{{\bf p}_i^2
+ m_i^2}$. The total 4-momentum is conserved,
$P^\mu \equiv {p_1'}^\mu + {p_2'}^\mu \equiv \sum_{i=1}^N p_i^\mu$.
Up to constants depending on ${p_1'}^\mu$ and ${p_2'}^\mu$, the total
cross section for this process is \cite{hagedorn}
\begin{equation}
\sigma
\sim \int \prod_{i=1}^N \, \frac{{\rm d}^3{\bf p}_i}{E_i} \;
\delta^{(4)}\left(P^\mu-\sum_{i=1}^N p_i^\mu\right)\; 
|{\cal M}(P,p_1, \ldots ,p_N)|^2 \,\, ,
\end{equation}
where the integration is over the $3\,N$ dimensional momentum space
of particles in the final state, the delta function
represents energy conservation, and $|{\cal M}|^2$ is the
modulus squared of the matrix element for the $2 \rightarrow N$ process.
The single-inclusive, invariant cross section 
for the production of a certain particle species -- without loss of generality
assumed to have 4-momentum $p_1^\mu$ --
is then (again, up to constants)
\begin{equation}
E_1 \, \frac{{\rm d} \sigma}{{\rm d}^3 {\bf p}_1}
\sim \int \prod_{i=2}^N \, \frac{{\rm d}^3{\bf p}_i}{E_i} \;
\delta^{(4)}\left(P^\mu-p_1^\mu-\sum_{i=2}^N p_i^\mu\right)\; 
|{\cal M}(P,p_1, \ldots ,p_N)|^2 \,\, .
\end{equation}
Let us introduce the ``average'' matrix element,
\begin{eqnarray}
\lefteqn{
\left\langle |{\cal M}(P,p_1, \ldots ,p_N)|^2 \right\rangle_{p_2,\ldots,p_N}}
\nonumber \\
& \equiv & \left[ \Phi(P-p_1) \right]^{-1}\; 
\int \prod_{i=2}^N \, \frac{{\rm d}^3{\bf p}_i}{E_i} \;
\delta^{(4)}\left(P^\mu-p_1^\mu-\sum_{i=2}^N p_i^\mu\right)\; 
|{\cal M}(P,p_1, \ldots ,p_N)|^2\,\, ,
\end{eqnarray}
where the average is over the $3\, (N-1)$ dimensional 
momentum space of the unobserved $N-1$ particles in the 
single-inclusive production of the
particle with momentum $p_1$, and
\begin{equation}
\Phi(P-p_1) \equiv \int \prod_{i=2}^N \, \frac{{\rm d}^3{\bf p}_i}{E_i} \;
\delta^{(4)}\left(P^\mu-p_1^\mu-\sum_{i=2}^N p_i^\mu\right)
\end{equation}
is the corresponding Lorentz-invariant momentum-space volume.
The momentum-space volume has dimension MeV$^{2(N-3)}$.
Then, the single-inclusive cross section can be written as
\begin{equation} \label{singinc}
E_1 \, \frac{{\rm d} \sigma}{{\rm d}^3 {\bf p}_1}
\sim \Phi(P-p_1) \; \left\langle
|{\cal M}(P,p_1, \ldots ,p_N)|^2 \right\rangle_{p_2,\ldots,p_N}\,\, .
\end{equation}
Dynamical information about the $2 \rightarrow N$ process is
contained in the second term only. The first factor, the
momentum-space volume, is in some sense trivial. 

To compute the momentum-space volume, for the sake of simplicity
(and because it allows us to obtain a purely analytic result)
let us assume that all $N$ particles in the final state are massless.
In this case, no other scale with the dimension of energy enters
$\Phi(P-p_1)$, and for dimensional reasons,
\begin{equation}
\Phi(P-p_1) \sim \left[(P-p_1)^2 \right]^{N-3}\,\, .
\end{equation}
Because $\Phi(P-p_1)$ is Lorentz-invariant, one may evaluate the
right-hand side in the C.M.\ frame of the incoming particles,
where $P^\mu \equiv (E,{\bf 0})$,
with the result
\begin{equation}
\Phi(P-p_1) \sim E^{2(N-3)} \left( 1 - \frac{ E_1}{E/2N}\, \frac{N-3}{N}\,
\frac{1}{N-3} \right)^{N-3}\,\, .
\end{equation}
In the limit $N \gg 1$, the term in parentheses is a representation
for the exponential function, 
$\lim_{n \rightarrow \infty} (1+x/n)^n \equiv e^x$,
such that
\begin{equation}
\Phi(P-p_1) \sim \exp \left( - \frac{E_1}{E/2N} \right)
\,\, , \;\;\;\; N \gg 1\,\, .
\end{equation}
Now insert this into Eq.\ (\ref{singinc}), and divide by the
total cross section to obtain the invariant momentum spectrum.
Then, for a given rapidity $y_1$, for instance $y_1 = 0$, 
one obtains
\begin{equation} \label{result1}
\left .\frac{{\rm d} N}{{\rm d} y_1\, {\rm d}^2 {\bf p}_{T,1}}
\right|_{y_1 = 0}
\sim \exp\left( - \frac{p_{T,1}}{E/2N} \right)\,\, .
\end{equation}
Provided that the average matrix element squared 
$\langle |{\cal M}|^2 \rangle$ does not contain a strong (exponential)
dependence on $p_{T,1}$, the invariant momentum spectrum decreases
exponentially with $p_{T,1}$, with an inverse slope parameter
$T_{\rm slope} = E/2N$, which proves our original assertion. 
The exponential behavior of the transverse momentum spectra
is due to phase space \cite{hormuz}.
No assumption about thermal equilibration is necessary to
obtain this result.

Nevertheless, single-inclusive particle spectra are
also exponential in thermal equilibrium, 
{\it i.e.}, for a system at temperature $T$. In this case, 
the invariant transverse momentum spectrum is 
given by the Cooper-Frye formula \cite{cooper}
\begin{equation} \label{cooperfryeformula}
E_1\, \frac{{\rm d}N}{{\rm d}^3 {\bf p}_1}
= \int_\Sigma {\rm d} \Sigma \cdot p_1 \;f\left(\frac{p_1 \cdot u}{T},
\lambda_1 \right)\,\, ,
\end{equation}
where $\Sigma$ is the 3-dimensional space-time hypersurface
on which the transverse momentum spectrum is
computed, ${\rm d} \Sigma^\mu$ is the normal vector
on $\Sigma$, and $f(x,\lambda)$ is the thermal distribution
function. 
For Boltzmann particles,
$f(x,\lambda) \sim \lambda\, \exp(-x)$; $\lambda \equiv
\exp(\mu/T)$ is
the fugacity of the particle, and $\mu$ the chemical potential.
The 4-vector $u^\mu$ in Eq.\ (\ref{cooperfryeformula})
is the 4-velocity of the system, {\it i.e.}, the
{\em average\/} 4-velocity of {\em particle flow}. In the rest frame
of the system, $u^\mu \equiv (1, {\bf 0})$. 

For the sake of simplicity, 
compute the spectrum at constant
time, where ${\rm d} \Sigma^\mu
= ({\rm d}^3 {\bf x}, {\bf 0})$, such that
\begin{equation}
E_1\, \frac{{\rm d}N}{{\rm d}^3 {\bf p}_1}
\sim V\, \lambda_1 \, E_1 \, \;\exp \left(- \frac{E_1}{T} \right) \,\, .
\end{equation}
For ultrarelativistic particles, the energy per particle is
related to the temperature via $E/N = 3\, T$, such that at midrapidity
\begin{equation} \label{result2}
\left. \frac{{\rm d}N}{{\rm d}y_1 \, {\rm d}^2 {\bf p}_{T,1}}
\right|_{y_1=0}
\sim \exp \left(- \frac{p_{T,1}}{E/3N} \right) \,\, .
\end{equation}
This is rather similar to Eq.\ (\ref{result1}), except that
the (inverse) slope of the $p_T$-spectrum is equal to the true thermodynamic
temperature $T= E/3N$, while previously, $T_{\rm slope} 
= E/2N$ is a factor 3/2 larger. 

The origin of this discrepancy is that invariant momentum space is
not identical to thermodynamic momentum space. For a single particle,
the former is ${\rm d}^3 {\bf p}/E$, while the latter is
${\rm d}^3{\bf p}$. The missing factor of $1/E$ is responsible for
the difference in the exponential slope. 

From an experimental point of view, without a fully
exclusive measurement one cannot precisely tell the number of particles
in the final state, such that the energy per particle $E/N$ is unknown. 
The important point is that then
{\em there is no possibility to distinguish between the two cases\/}
Eqs.\ (\ref{result1}) and (\ref{result2}): usually, the strategy is
to perform an exponential fit to transverse momentum spectra with 
a slope parameter $T_{\rm slope}$, but in this way one cannot
test whether the relationship between $E/N$ and
$T_{\rm slope}$ is the same as in thermal equilibrium.
Although the spectra are exponential,
like for a thermal system, the system need not be in thermal
equilibrium. Although the slope parameter is
a quantity with the same dimension as temperature, 
temperature is not defined, if the system is not in
thermal equilibrium. One may even go one step further:
although single-inclusive
spectra for {\em different\/} particle species may have the same
slope parameter, this does not mean that the system is in thermal
equilibrium. It is simply due to the fact that the slope is
proportional to the energy per particle $E/N$, which is a constant
for all $N$ particles in the final state of 
a $2 \rightarrow N$ process at a given C.M.\ energy $E$.

\section{KINETIC FREEZE-OUT} \label{III}

Slope parameters for the most abundant particle species show different
behavior as a function of particle mass in $AA$
as compared to $pp$ collisions \cite{NA49}. In $AA$
collisions, they increase linearly with the particle mass, 
\begin{equation}
T_{{\rm slope},\, AA} \simeq a + b\; m \,\, ,
\end{equation}
while in $pp$ collisions, they are independent of the particle mass,
\begin{equation}
T_{{\rm slope},\, pp} \simeq a \,\, .
\end{equation}
For CERN-SPS energies, $a \simeq 140$ MeV for both $AA$
and $pp$ collisions, while $b$ depends on the
system size; it increases with the mass number $A$ of the nuclei.

Although there might be other explanations for this behavior, the most natural
interpretation is in terms of collective motion. The parameter $a$
is the slope parameter determined by multiparticle momentum space.
Since the energy per particle $E/N$ is an intensive quantity, 
in both cases $a \sim E/N$ is independent of the system size.
Once the particles are created, however, the environment matters, {\it i.e.},
whether there are many final-state interactions, like in $AA$
collisions, or whether the particles more or less
freely stream towards the detector, like in $e^+e^-$ and $pp$ collisions. 
As explained in the introduction,
in the first case local kinetic equilibration of the system
leads to pressure gradients, which in turn generate collective motion.
The constant $b$ parametrizes this collective motion; it is
proportional to the (average) collective flow velocity of the expanding 
system \cite{csorgo}. 
The larger the system, the more final-state interactions occur, and
the larger is the collective flow velocity. Consequently, 
$b$ increases with the system size.

A word of caution is in order. The fact that there is
collective motion in $AA$ collisions does not necessarily 
mean that local kinetic equilibrium is established and,
consequently, ideal fluid dynamics applies to determine the
evolution of the system.
Pressure gradients which drive collective motion occur also in 
systems away from equilibrium.
Nevertheless, the fact that there is collective motion indicates
that there are final-state interactions, which will {\em eventually\/} 
drive the  system towards kinetic equilibrium, unless the 
macroscopic expansion rate considerably exceeds the microscopic
scattering rate. Therefore, the ideal-fluid approximation for
the dynamical evolution of the system might not be too bad, especially
for collisions of large nuclei at ultrarelativistic energies.

Freeze-out of particles occurs when
microscopic interaction rates become small as compared to
the macroscopic collective expansion rate. 
In fluid dynamical models, 
one commonly assumes kinetic freeze-out to happen instantaneously
along space-time hypersurfaces of constant density or temperature.
This is certainly an idealization: microscopically,
a particle has a certain probability to decouple anywhere and anytime
during the evolution of the system \cite{bass}.

The invariant momentum spectra of frozen-out particles are computed according
to Eq.\ (\ref{cooperfryeformula}), with
the freeze-out hypersurface $\Sigma \equiv \Sigma_{\rm f.o.}$.
There are a number of conceptual problems with 
this formula, a discussion of which is beyond the scope of
this talk, see Refs.\ \cite{bugaev} for more details. 
In practice, however, the Cooper-Frye formula (\ref{cooperfryeformula})
is sufficient to compute the particle spectra to
reasonable accuracy. 

Let us assume that kinetic
freeze-out happens across a surface of constant temperature $T_{\rm f.o.}$.
Applying the mean-value theorem to Eq.\ (\ref{cooperfryeformula}),
the spectrum of frozen-out particles at midrapidity $y=0$ is
\begin{equation} \label{spectrum}
\left. \frac{{\rm d} N}{{\rm d} y\,{\rm d}^2 {\bf p}_T} \right|_{y=0}
\sim \exp \left( -\langle \gamma \rangle \frac{ m_T - 
{\bf p}_T \cdot \langle {\bf v}_T \rangle }{T_{\rm f.o.}} \right) \,\, ,
\end{equation}
where $\langle {\bf v}_T \rangle$ and $\langle \gamma \rangle$
are suitably defined {\em average\/} values for the transverse
fluid 3-velocity and the Lorentz gamma factor
along the freeze-out hypersurface. At $y=0$, it is reasonable
to neglect longitudinal collective motion, such that $\gamma 
\simeq (1- {\bf v_T}^2)^{-1/2}$ is determined by ${\bf v}_T$.
Assuming azimuthal symmetry, the spectrum (\ref{spectrum}) then
depends only on two parameters, the kinetic 
freeze-out temperature $T_{\rm f.o.}$
and the modulus of the average transverse collective flow velocity 
$\langle v_T \rangle$.

\begin{table}[ht]
\caption{Kinetic freeze-out temperatures and average
transverse flow velocities for $PbPb$ collisions at
$E_{\rm Lab} = 158\, A$GeV.}
\label{table1}
\renewcommand{\tabcolsep}{2pc} % enlarge column spacing
\renewcommand{\arraystretch}{1.2} % enlarge line spacing
\begin{tabular}{@{}lcc}
\hline
Ref.           & $T_{\rm f.o.}$ [MeV] & $\langle v_T \rangle$     \\
\hline
\cite{schlei}  &    140              &    0.55                    \\
\cite{hung}    &   110 -- 120        &    0.60                    \\
\cite{huovinen}&   120 -- 140        &                            \\
\cite{dumitru} &   130               &    0.50                    \\
\cite{kampfer} &   120               &    0.43                    \\
\cite{tomasik} &   100               &    0.55                    \\
\cite{ster}    &   140               &    0.55                    \\
\hline
\end{tabular}\\[2pt]
\end{table}

Table \ref{table1} shows values for $T_{\rm f.o.}$ and $\langle v_T \rangle$
extracted by several authors
from experimental spectra for $PbPb$ collisions at
CERN-SPS energies, $E_{\rm Lab} = 158\, A$GeV.
From Eq.\ (\ref{spectrum}) it is obvious that $T_{\rm f.o.}$
and $\langle v_T \rangle$ are correlated:
fits of similar quality can be obtained
by trading off a lower value for $T_{\rm f.o.}$ against
a higher value for $\langle v_T \rangle$ and vice versa. This
can also be seen in the values quoted in Table \ref{table1}.
This ambiguity can be removed by two-particle correlations,
where $T_{\rm f.o.}$ and $\langle v_T \rangle$ are
correlated in the opposite way \cite{tomasik}.

\section{CHEMICAL FREEZE-OUT} \label{IV}

In principle, there is another, independent way to determine 
thermodynamic quantities at freeze-out. 
First note that, in order to compute the 
transverse momentum spectrum (\ref{spectrum}) of particle species
$i$, one not only needs to know the
temperature and the transverse flow velocity along the
freeze-out hypersurface,
which determine the shape of the spectrum as a function
of $p_T$, but also the fugacity
$\lambda_{i, {\rm f.o.}} \equiv \exp (\mu_{i, {\rm f.o.}}/T_{\rm f.o.})$, 
which determines the 
absolute normalization of the spectrum, cf.\ Eq.\ (\ref{cooperfryeformula}). 
In general, $\mu_{i, {\rm f.o.}}$ varies along the freeze-out hypersurface, 
but in order to proceed, let us make the assumption (the first of three)
that it is constant, like
the temperature $T_{\rm f.o.}$. In the following, we shall
drop the subscript ``f.o.'', but remember that all
thermodynamic quantities, as well as the flow 4-velocity $u^\mu$,
are taken on the freeze-out hypersurface.

The second assumption we shall make is that {\em all\/} particle
species freeze out across the {\em same\/} freeze-out hypersurface.
This is certainly an idealization, as the mean free paths of
different particle species are different \cite{bass}.
Under these two assumptions, the ratio of the {\em total\/}
particle numbers of particle species $i$ and $j$ is
\begin{equation} \label{ratio}
\frac{N_i}{N_j} \equiv \frac{ \int ({\rm d}^3{\bf p}_i/E_i)\,
\int_\Sigma {\rm d} \Sigma \cdot p_i\, f(p_i \cdot u/T, \lambda_i)}{
\int ({\rm d}^3{\bf p}_j/E_j)\,
\int_\Sigma {\rm d} \Sigma \cdot p_j\, f(p_j \cdot u/T, \lambda_j)}
\equiv \frac{\int_\Sigma {\rm d} \Sigma \cdot {\cal N}_i}{ 
\int_\Sigma {\rm d} \Sigma \cdot {\cal N}_j } \,\, ,
\end{equation}
where
\begin{equation}
{\cal N}_i^\mu \equiv \int \frac{{\rm d}^3 {\bf p}_i}{E_i} \,p_i^\mu
f\left( \frac{p_i \cdot u}{T}, \lambda_i \right) 
\end{equation}
is the 4-current of particle species $i$. In kinetic equilibrium
the 4-current assumes the simple form ${\cal N}_i^\mu \equiv
n_i\, u^\mu$, where $n_i = n_i(T,\mu_i)$ is the density
of particles of species $i$ in the local rest frame. 
The flow 4-velocity $u^\mu$ is common to all particle species
because the system is assumed to be kinetically equilibrated
immediately prior to freeze-out. Equation (\ref{ratio})
becomes \cite{cleymansredlich}
\begin{equation} \label{ratio2}
\frac{N_i}{N_j} \equiv \frac{n_i(T,\mu_i)}{n_j(T,\mu_j)} \,\, ,
\end{equation}
because the factor $\int_\Sigma {\rm d}\Sigma \cdot u$ cancels
between numerator and denominator. 
The result (\ref{ratio2}) is remarkable in the sense that, under
the present assumptions, the ratio of total particle yields 
is {\em independent\/} of the detailed dynamical evolution of
the system {\em prior\/} to freeze-out. Moreover, this ratio is the same as
for a system in {\em global\/} kinetic equilibrium at temperature $T$ with
chemical potentials $\mu_1, \, \mu_2, \ldots$.

Let us now make the third assumption, namely that
the system is not only in kinetic,
{\it i.e.}, thermal and mechanical equilibrium, but also
in {\em chemical\/} equilibrium. 
This assumption is fulfilled if inelastic (and not only elastic)
collision rates are sufficiently large.
In this case, the chemical potential of particle species $i$
is given by
\begin{equation} \label{mu}
\mu_i = b_i\; \mu_B + s_i\; \mu_S + e_i\; \mu_e + \ldots \,\, ,
\end{equation}
where $\mu_B,\, \mu_S,$ and $\mu_e$ are baryon, strangeness, and electric
charge chemical potentials, respectively, and $b_i,\, s_i, \, e_i$ are
baryon, strangeness, and electric charges of particle species $i$.
The ellipsis in Eq.\ (\ref{mu}) stands for possible other 
conserved charges with associated chemical potentials.

Global baryon, strangeness, and charge conservation impose
additional conditions which allow to eliminate all charge chemical
potentials except for one, {\it e.g.} $\mu_B$. Then, all particle
ratios are a function of two parameters only, $T$ and $\mu_B$.
Taking into account the known (vacuum) values for the hadronic
masses $m_i$ and the (vacuum) partial decay widths
$\Gamma_{i \rightarrow jk \ldots}$, one can then
extract $T$ and $\mu_B$ from ($4 \pi$ extrapolated)
data via a $\chi^2$ analysis.

Different authors use slightly different strategies to perform
such $\chi^2$ fits. Some introduce a hard-core repulsion between
hadrons \cite{PBM,gorenstein}, 
some relax the assumption of chemical equilibration
of strangeness \cite{becattini,rafelski}, some conserve
strangeness exactly in the canonical ensemble \cite{cleymans}.
In essence, all methods introduce at least one additional
parameter, which of course improves the quality of the fit, but
does not fundamentally change the underlying assumptions. 
All fits are roughly of the same quality, which makes it hard
to draw definite conclusions regarding the
necessity of the individual approach.

\begin{figure}[t]
%\vspace*{-2cm}
\begin{center}
\epsfig{file=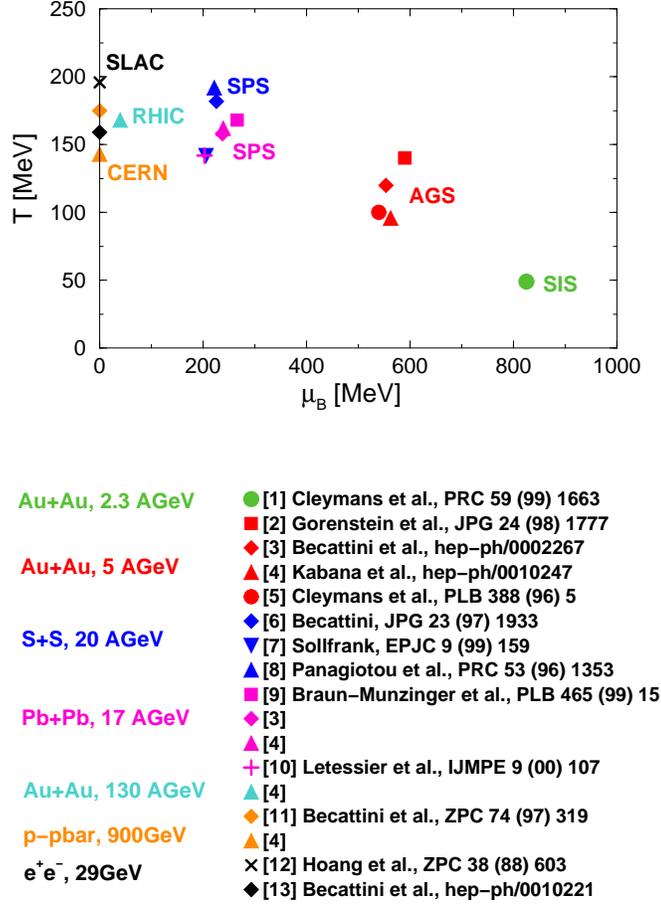,height=12cm}
\end{center}
\vspace*{-1cm}
\caption{Values for temperature $T$ and chemical potential $\mu_B$
extracted from particle ratios. Different collision energies and
systems are shown in different colors. Different symbols indicate differences 
in the fitting procedure, for more details see listed
references (numbering not identical with that in main text).}
\label{figTmu}
\end{figure}

Results of $\chi^2$ analyses for 
$e^+e^-$, $pp$, as well as $AA$ collisions
with equal-mass nuclei at various collision energies
are compiled in Fig.\ \ref{figTmu} (see also \cite{cleymansredlich,sollfrank}
for a similar compilation).
Two features are remarkable: first, as noted by Cleymans and Redlich
\cite{cleymansredlich}, all $(T,\mu_B)$ combinations fall
on a line of constant energy per particle, $\langle E \rangle / \langle
N \rangle \simeq 1$ GeV. The value 1 GeV 
can be intuitively understood as setting the
energy scale below which inelastic collisions cease and therefore chemical
equilibration becomes impossible.

The second feature is that, at CERN-SPS energies and above, the
freeze-out temperature is $T \sim 160$ MeV, which is somewhat higher than
the kinetic freeze-out temperature discussed in the last section.
This can be understood rather naturally noting that
the extraction of a freeze-out temperature from particle ratios assumed 
chemical equilibrium, while the extraction from single-inclusive
spectra only assumed kinetic equilibrium.
Thus, particle ratios determine the temperature at {\em chemical\/} freeze-out,
and not at kinetic freeze-out. Since chemical equilibrium requires 
frequent {\em inelastic\/} collisions, while kinetic
equilibrium only requires frequent {\em elastic\/} collisions,
chemical freeze-out occurs earlier, {\it i.e.}, at larger
temperatures, than kinetic freeze-out.
In order to distinguish quantities at chemical from those at 
kinetic freeze-out, in the following the former will 
be denoted with a subscript ``ch.'', and the latter with the subscript
``f.o.'', as above.

Given a set of hadronic masses $m_i$ and decay widths
$\Gamma_{i \rightarrow jk \ldots}$, 
the extracted values of $(T, \mu_B)_{\rm ch.}$ 
displayed in Fig.\ \ref{figTmu} are relatively insensitive
to errors in the experimental determination of
total particle numbers. This can be seen in the non-relativistic
approximation,
\begin{equation}
n_i(T, \mu_i) \sim \exp \left( \frac{\mu_i - m_i}{T} \right)\,\, ,
\end{equation}
such that Eq.\ (\ref{ratio2}) becomes
\begin{equation} \label{ratio3}
\frac{N_i}{N_j} \sim \exp \left( 
\frac{\mu_{i, {\rm ch.}}-\mu_{j, {\rm ch.}}}{T_{\rm ch.}} - 
\frac{m_i-m_j}{T_{\rm ch.}} \right)\,\,.
\end{equation}
Errors in the determination of $N_i$ and $N_j$ influence the fitted
values for $T_{\rm ch.}$ and $\mu_{B, {\rm ch.}}$
only logarithmically.

On the other hand, as can be also seen from Eq.\ (\ref{ratio3}),
the values for the hadronic masses $m_i$ entering
the fit influence $T_{\rm ch.}$ and $\mu_{B, {\rm ch.}}$
{\em linearly}.
Most fits assume that the hadronic masses and decay widths 
at chemical freeze-out are the same as in vacuum.
At such large temperatures 
and chemical potentials, however, a change
of mass and decay width
due to in-medium interactions
cannot be excluded \cite{brown,zschiesche,florkowski},
$m_i \rightarrow m_i^*,\; \Gamma_{i \rightarrow jk} \rightarrow
\Gamma^*_{i \rightarrow jk}$. These interactions are elastic collisions 
which still occur at chemical freeze-out, since the system remains
{\em kinetically\/} equilibrated. 
In \cite{zschiesche}, such an analysis was performed for $PbPb$ collisions
at CERN-SPS energies in the framework of a chiral model.
In this model, masses in general decrease at large
density and temperature, such that the value for the chemical freeze-out
temperature is smaller than for a fit with vacuum masses, $T_{\rm ch.} \simeq
144$ MeV.
Interestingly, in the calculation of \cite{zschiesche}
it turns out to be approximately the same as the {\em kinetic\/}
freeze-out temperature, although at kinetic freeze-out, particles
{\em have\/} to attain their vacuum masses, due to the absence
of any kind of interaction.
Clearly, more work has to be done to understand and
possibly refine the freeze-out picture,
for instance, also to include in-medium modifications of the
decay widths.

\section{CONCLUSIONS} \label{V}

In this talk, I first discussed single-inclusive, invariant particle 
spectra in a $2 \rightarrow N$ process. I explicitly demonstrated
that the momentum space volume of the final-state particles gives
rise to the exponential nature of these spectra. For the special case of
massless particles, I showed that
the slope of the spectra differs from the one for a thermally equilibrated
system of particles, {\it i.e.}, that
exponential particle spectra alone are {\em not\/} indicative for
the existence of thermal equilibrium.

I then argued that the slope characteristics for different particle
species suggest collective motion in $AA$ collisions.
Single-inclusive, invariant particle spectra then yield
information about the {\em average\/} temperature $T_{\rm f.o.}$ and 
{\em average\/} collective flow velocity $\langle v_T \rangle$
at kinetic freeze-out. 

Finally, I discussed the extraction of the chemical freeze-out parameters
$T_{\rm ch.}$ and $\mu_{B, {\rm ch.}}$
from particle ratios. This analysis is based upon rather
restrictive assumptions, and depends sensitively on the hadronic
masses and decay widths.

Nevertheless, it is astonishing that a fit with 
essentially two parameters
can reproduce a multitude of data with reasonably good quality.
There could be several explanations. First, the state preceding
chemical freeze-out is indeed in thermodynamic equilibrium.
The degrees of freedom could be hadrons, but
since $T_{\rm ch.}$ is close to the temperature of the 
confinement-deconfinement transition, it is, however, also conceivable that 
at some earlier stage in the evolution of the system
a thermodynamically equilibrated QGP has existed \cite{PBM,cern}.
From thermodynamic arguments alone, however, one can never
distinguish between these two possibilities: by definition, a state 
in thermodynamic equilibrium has no knowledge about its past. 

The discussion of section \ref{II} allows for a second explanation.
Multiparticle production saturates the available phase space, such that
final-state hadrons have exponential spectra with a slope that
can be interpreted as a temperature, and an absolute normalization
that can be interpreted as a fugacity.
Although the discussion of section \ref{II} has shown that the
final state need not be thermodynamically equilibrated to have
these properties, it then at least looks like hadrons
are ``born into thermodynamic equilibrium''
\cite{stock}. In principle, the chemical composition 
could then immediately freeze out. In this case, no additional 
assumptions about
an equilibrated state {\em preceding\/} chemical freeze-out are necessary. 
As seen in section \ref{III}, there
certainly are elastic interactions which cause collective motion, but
this could in principle happen after chemical freeze-out.

To distinguish between the first and second scenario,
a more thorough understanding
of multiparticle production in $AA$ collisions is necessary.
The only way to achieve this is to compare $AA$
with $pp$ and $pA$ collisions. It is therefore mandatory to gather
more experimental data on the latter at comparable collision energies.

\subsection*{Acknowledgements} 

I am indebted to J.\ Cleymans, B.\ Cole,
K.\ Kajantie, R.\ Kuhn, K.\ Redlich, and E.\ Shuryak for
discussions. Thanks go 
Columbia University's Nuclear Theory Group for
providing the computing facilities necessary to complete this work.

\end{document}